\documentclass{aa}
\usepackage{graphicx} 
\begin{document}
   \thesaurus{
              (02.02.1; 
11.01.2; 
		11.14.1; 
11.17.2; 
11.17.3 
		)} 
%
\newcommand{\sy}{{Seyfert }}
\newcommand{\etal}{{ et al. }}
\newcommand{\beq}{\begin{equation}}
\newcommand{\eeq}{\end{equation}}
\newcommand{\msun}{M_{\odot }}

\title{Black hole mass estimation with a relation between the BLR size and 
emission line luminosity of AGN}
\author{Xue-Bing Wu$^1$,  R. Wang$^1$, M.Z. Kong$^2$, F.K. Liu$^1$, J.L. Han$^2$} 
\institute{$^1$Department of Astronomy, Peking University, Beijing 100871, China\\
$^2$National Astronomical Observatories of Chinese Academy of Sciences, Beijing 100012, China
}

\date{}
\offprints{wuxb@bac.pku.edu.cn}

\authorrunning{Wu et al.}
\titlerunning{Black hole mass estimation of AGN}
\maketitle

\begin{abstract}
 
An empirical relation between the broad line region (BLR) size and  
optical continuum luminosity
is often adopted to estimate the BLR size and then the black hole mass of AGNs. 
However,  optical luminosity may not be
a good indicator of photoionizing luminosity  for extremely 
radio-loud AGNs
 because the jets usually contribute significantly
to the optical continuum. Therefore, the black hole masses 
derived for blazar-type AGNs with this method are probably overestimated. 
Here we first derived a tight
empirical relation between the BLR size and the H$_\beta$ emission line luminosity, 
$R(\rm{light-days})= 24.05(L_{H_\beta}/10^{42} ergs~s^{-1})^{0.68}$,
from a sample of 34 AGNs with the BLR size estimated with the reverberation 
mapping technique. Then we applied this relation to estimate the black hole masses of
some AGNs and found that for many extremely radio-loud AGNs the black hole masses 
obtained with the $R-L_{H_\beta}$ relation are systematically lower
than those derived previously with the $R-L_{5100\AA}$ relation, while 
for radio-quiet and slightly radio-loud AGNs the results obtained with these two methods 
are almost the same. The difference of black hole masses
estimated with these two relations increases with the radio-loudness for extremely 
radio-loud AGNS, 
which is consistent with the fact that their equivalent widths of H$_{\beta}$
emission line become smaller at higher radio-loudness.
If the small H$_{\beta}$ equivalent widths of extremely radio-loud
 AGNs are indeed caused by the beaming effect, we argue that the optical emission 
line luminosity may be a better tracer of ionizing luminosity for  blazar-type AGNs 
and the black hole mass derived with the $R-L_{H_\beta}$ relation  are probably 
more accurate.


\end{abstract}

\section{Introduction}
Supermassive black hole is essential  for 
AGNs activities (Lynden-Bell 1969; Rees 1984).
The black hole masses of some nearby AGNs have been recently 
estimated by the reverberation mapping technique (Wandel, Peterson \& Malkan 1999; 
Ho 1999; Kaspi \etal 2000), with which the size 
of the broad line region (BLR) can be measured from the time delay between the 
flux variations of the continuum and the emission lines of AGNs. The black hole mass is
then estimated using the Virial theorem from the BLR size and the characteristic velocity 
(determined by the full width at half-maximum (FWHM)
of emission line). So far, the reverberation studies have
yielded the black hole masses of about 20 
Seyfert 1 galaxies and 17 nearby bright quasars. 

An empirical relation between the 
BLR size ($R$) and the optical continuum luminosity at 5100$\AA$ ($L_{5100\AA}$) has been 
derived by Kaspi et al. (2000) using the observed data of 34 nearby AGNs. 
Because the  measurement of the BLR size with the reverberation mapping technique
needs long-term monitoring of continuum and emission line fluxes, it is  
impractical for most AGNs. Therefore, the empirical relation has been frequently adopted
to estimate the BLR size and then
derive the black hole masses  for AGNs in some samples of mostly radio-quiet objects 
(Laor 2000; McLure \& Dunlop 2001; Wandel 2002),
and of purely radio-loud objects (Lacy et al. 2001; Gu, Cao \& Jiang 2001; Oshlack, 
Webster \& Whiting 2002). 
However, the optical luminosity of  some radio-loud AGNs  (especially blazars), may not
be a good indicator of ionizing luminosity, which is usually related to the UV/optical
radiation from the accretion disk around the central black hole. The relativistic jets of
 blazar-type AGNs not only dominate the radio and high energy X-ray and $\gamma$ 
ray 
radiation, but also significantly contribute to the optical luminosity 
 in some cases (Scarpa \& Urry 2002).  For example, 
many optical jets
have been discovered recently in AGNs by the HST (Scarpa et al. 1999; Jester 2003; 
Parma \etal 2003), which clearly
suggests that the jets contribute significantly in the optical band. Furthermore, optical 
synchrotron radiation
has been  detected for many other radio-loud AGNs (Whiting, Webster \& Francis 2001;
Chiaberge, Capetti, \& 
Celloti 2002; Cheung \etal 2003). 
Therefore, the measured optical continuum luminosity of  some  extremely radio-loud 
AGNs is significantly contributed by
the optical radiation from the jets and may be much larger than the ionizing luminosity
required to produce broad emission lines. 
Using the empirical relation between the BLR
size and optical luminosity at 5100$\AA$, which was obtained based on the sample of 
mostly radio-quiet AGNs (Kaspi \etal 2000), one would significantly overestimate the actual
BLR size and hence the black hole mass of these radio-loud AGNs. Oshlack et al. (2002) 
have shown that their estimated black hole
masses would be lower if the synchrotron contribution to the optical flux is subtracted.
However, it is not easy to make such a correction for a large sample of radio-loud AGNs.
In addition, the contribution of the host galaxy to the
optical continuum should also be taken into account especially when the host galaxy of 
AGNs can be resolved optically.
Therefore, optical luminosity may not be a good indicator of photoionization luminosity  
of AGNs in some cases. 

In this paper we will first derive an empirical relation between the BLR size and 
the H$_\beta$ emission line luminosity for 34 AGNs in the sample of reverberation mapping
studies. We  then argue that the BLR size obtained from the H$_\beta$ luminosity is more
reasonable at least for  some extremely radio-loud AGNs. Finally we  apply this new 
empirical relation to estimate
the black hole masses of some quasars and compare them with previous results.

\section{The relation between the BLR size and H$_\beta$ luminosity}
Kaspi et al. (2000) have compiled the observational data of 17 Seyfert galaxies 
(Wandel, Peterson \& Malkan 1999) 
and 
17 nearby quasars with black hole masses estimated with the reverberation mapping 
technique.  Using a linear fit to the available data with errors, they got an 
empirical relation between the 
BLR size and the optical
continuum luminosity at 5100$\AA$ as:
$$
R_{BLR}(\rm{light-days})=(32.9^{+2.0}_{-1.9})~~~~~~~~~~~~~~~~~~~~~~
$$
\beq
~~~~~~~~~~~~~~~~~~~~~~~~~~[L_{5100\AA}/10^{44}ergs~s^{-1}]^{0.700\pm0.033}
 .
\eeq
 With an ordinary least square (OLS) 
bisector method (Isobe et al. 1990), we can obtain such a relation as
$R_{BLR}(\rm{light-days})=31.1[L_{5100\AA}/10^{44}ergs~s^{-1}]^{0.701}$, which is almost
the same as that shown above.
We should keep in mind that this $R-L_{5100\AA}$ relation was obtained with mostly radio-quiet 
AGNs. The optical continuum luminosity may
not be a good indicator for photoionizing luminosity for  some extremely radio-loud AGNs. 
Instead, the emission line luminosity may be a better indicator because it is free
from the beaming effects of the jet. 

Using the available data of BLR sizes 
and H$_\beta$ fluxes for 34 AGNs in the reverberation mapping studies, we can 
investigate the relation between the BLR size and the
H$_\beta$ emission line luminosity (including both broad and narrow components). 
 In Table 1 we listed the BLR size and the luminosity data of these 34 AGNs.
The H$_\beta$ luminosity is calculated from the H$_\beta$ flux which is available 
for  16 PG quasars,
8 AGNs in Ohio sample, and other 9 Seyfert 1 galaxies
 (see references listed in Table 1).
Because there is no available data of H$_\beta$ flux
for PG 1351+640 (Kaspi \etal 2000), we exclude this object from our investigation.
In addition, we add another Seyfert galaxy Mrk 279  in our sample because both the BLR
size and the H$_\beta$ flux have been measured recently (Santos-Lleo \etal 2001). 
All H$_\beta$ luminosity data  have been corrected for  
Galactic extinction using the values from NED\footnote{http://nedwww.ipac.caltech.edu}
 (see also Burstein \& Heiles 1982). 
The cosmology with
Hubble constant $H_0=75 km~s^{-1}Mpc^{-1}$ and deceleration parameter $q_0 =0.5$ were 
adopted through out the paper.

\begin{figure}
\includegraphics[width=9.5cm, height=12cm]{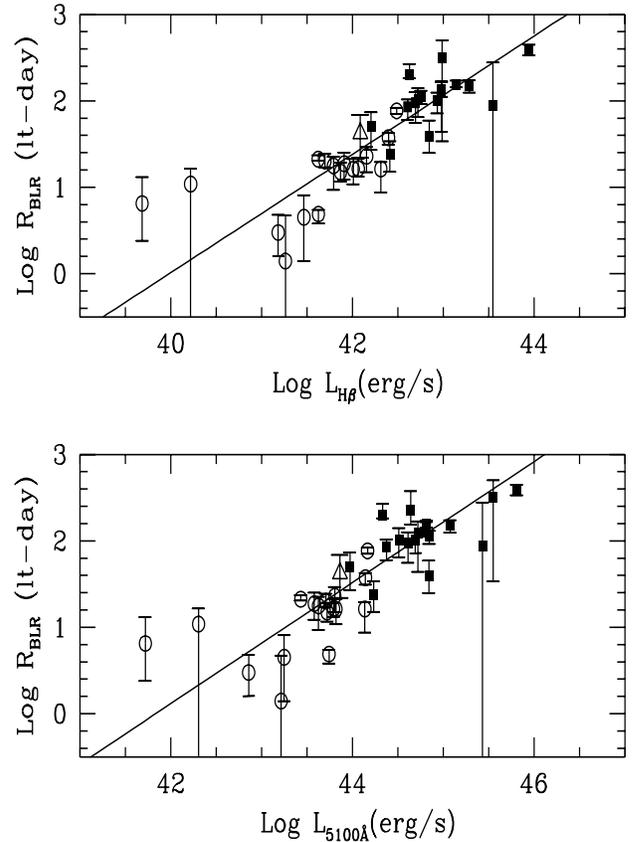}
\caption{The relations of BLR size and luminosity for 34 AGNs in the
reverberation mapping studies. The open and filled symbols denote Seyfert galaxies and
quasars respectively. The upper panel shows
the $R-L_{H_\beta}$ relation. The line shows the OLS bisector fit to the data.
 The lower panel shows the $R-L_{5100\AA}$ 
relation, which is 
identical to the Figure 6 in Kaspi \etal (2000) except that we include Mrk 279 (shown as 
open triangle). The line
represents the linear fit given in Kaspi \etal (2000).}
\end{figure}

With these data, we derive an empirical
relation between the BLR size 
and H$_\beta$ luminosity. 
With the OLS 
bisector method we obtained:
$$
\rm{Log}~R~(\rm{light-days}) = (1.381\pm0.080)+ ~~~~~~~~~~~~~~~~~~~~~~
$$
\beq
~~~~~~~~~~~~~~~(0.684\pm0.106) Log~(L_{H_\beta}/10^{42}~ergs~s^{-1}) .
\eeq
The slope of this relation is slightly flatter than that of $R-L_{5100\AA}$ relation given
in Kaspi \etal (2000),  consistent with $L_{H_\beta}\propto L_B^{0.93}$  
obtained by Ho \& Peng (2001) for PG quasars.
The Spearman's rank correlation coefficient of our $R-L_{H_\beta}$ relation is 0.91, slightly
higher than 0.83 for the $R-L_{5100\AA}$ relation (Kaspi \etal 2000), 
which implies that
the $R-L_{H_\beta}$ relation is slightly tighter than the $R-L_{5100\AA}$ relation.  
 We also used the linear fit to the data with errors that was used by Kaspi et al. (2000)
to derive the $R-L_{5100\AA}$ relation. Applying this method to the $R-L_{H_\beta}$ relation 
yields the same result as the OLS bisector method.

In Fig. 1 we show the dependence of the BLR size on $L_{H_\beta}$ and $L_{5100\AA}$. 
 The two relations are similar and thus the $R-L_{H_{\beta}}$ relation  can 
be an alternative of the  $R-L_{5100\AA}$ relation in estimating the BLR size for 
{\it radio-quiet} AGNs.

\section{Comparison of black hole mass estimation of AGNs from two relations}
Since the $R-L_{5100\AA}$ relation obtained by Kaspi \etal (2000) has been frequently 
used to estimate the BLR size and the black hole mass of both radio-quiet and 
{\it radio-loud}
 AGNs, it is important to investigate the applicability of such an approach 
for radio-loud objects.

\begin{figure}
\includegraphics[width=9.5cm, height=12cm]{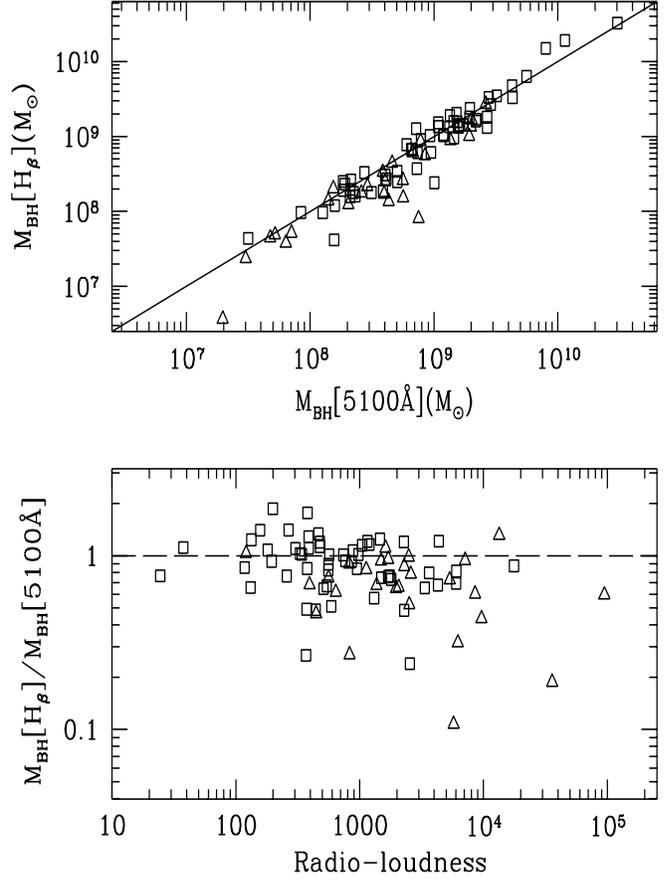}
 \caption{Upper panel: Comparison of the black hole masses of radio-loud quasars 
estimated with $R-L_{H_\beta}$ and 
$R-L_{5100\AA}$ relations. The squares represent 59 quasars in Brotherton (1996)
and the triangles represent 27 PHFS quasars in Oshlack \etal (2001).  The diagonal
line shows the relation where both masses are identical. Lower panel: The ratios of 
black hole masses estimated with two different 
relations are plotted against
 the radio-loudness of radio-loud quasars.  The dashed line indicates the case 
where the two black hole mass estimates are identical. Evidently $M_{BH}[5100\AA]$  becomes 
systematically larger than  $M_{BH}[H_\beta]$ at higher radio-loudness.}
\end{figure}

Brotherton (1996) studied the emission line properties of 59 radio-loud quasars. We 
adopted his published values of absolute V-band magnitude, equivalent width and FWHM of 
$H_{\beta}$ emission line. The
continuum luminosity at 5100$\AA$ and the $H_{\beta}$ luminosity (scaled to our 
cosmology parameters)
were calculated after considering the Galactic extinction and K-correction (optical 
spectral index was 
assumed to be 0.3). We then estimated the BLR size using both $R-L_{H_\beta}$ and 
$R-L_{5100\AA}$ relations and derived the black hole mass with the formula
$M_{BH}=3V_{FWHM}^2 R/4G$ (here we assumed the BLR velocity 
$V\sim (\sqrt{3}/2)V_{FWHM}$  as in Kaspi \etal 2000). With these two relations, we also 
estimated the black hole masses of another 27 radio-loud quasars with available
data of both the equivalent width and FWHM of $H_{\beta}$ emission line in the Parkes 
Half-Jansky flat-spectrum Sample (PHFS) (Drinkwater \etal 1997; Francis, Whiting \& Webster 
2000; Oshlack \etal 2001). We compared the black hole masses obtained with the 
$R-L_{H_\beta}$ and 
$R-L_{5100\AA}$ relations in Figure 2. Evidently the masses obtained with
the  $R-L_{H_\beta}$ relation are systematically lower that those obtained with
the $R-L_{5100\AA}$ relation for  some extremely radio-loud quasars.

In Figure 2 we also show how the difference of black hole masses obtained with these 
two relations varies with
the radio-loudness for these two samples of radio-loud quasars. The values of
radio-loudness of 27 PHFS
quasars were taken from Oshlack \etal (2001) and those of 59 quasars in  Brotherton
(1996) were calculated from the $R_V$ value (defined as the ratio of core radio
luminosity and the V-band optical luminosity) and the core and extended radio 
luminosity values listed in his Table 1. It is clear that 
the difference of  black hole masses  is small when the radio-loudness is small but becomes
larger as the radio-loudness increases.  The Spearman's rank correlation coefficient is
-0.34, implying a modest correlation between the difference of  black hole masses and
 the radio-loudness. For some  individual quasars with 
higher radio-loudness, the black hole mass estimated with the  $R-L_{5100\AA}$ 
relation can be 3$\sim$10 times larger than that estimated with the 
$R-L_{H_\beta}$ relation.  
 In Figure 3 we also plotted the equivalent width (EW) of H$_\beta$ emission line 
against the radio-loudness for objects in these two radio-loud AGN samples. We can see that the 
EW(H$_\beta$) becomes smaller at higher radio-loudness. Such a relation, although with only a 
modest 
 Spearman's rank correlation coefficient of -0.33,   indicates that
the smaller EW(H$_\beta$) of some extremely radio-loud AGNs could be  at least partly
due to 
the beaming
effects. 

\begin{figure}
\includegraphics[width=9.5cm, height=10cm]{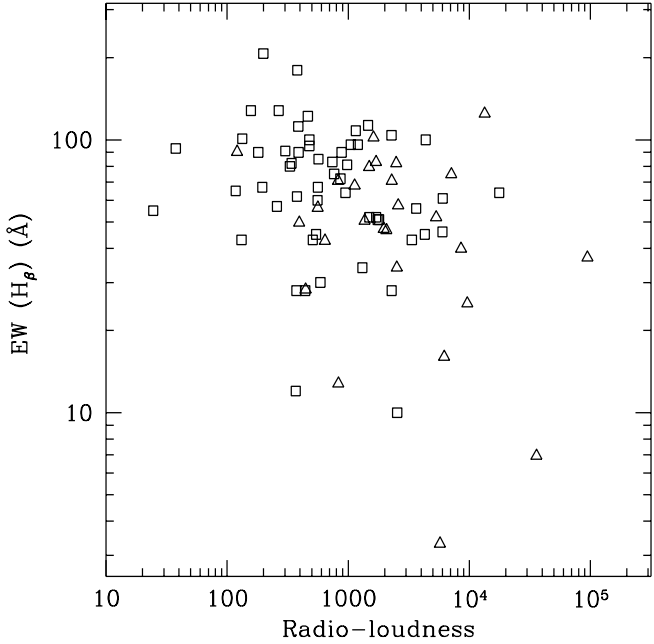}
 \caption{The relation between the equivalent width of H$_\beta$ emission line 
and the radio-loudness of radio-loud quasars. The symbols have the same meaning as in Figure 2.}
\end{figure}

For radio-quiet AGNs,  however, both the optical continuum 
and emission line luminosities are probably free from the jet 
contributions and  therefore both can be good tracers of photoionization 
luminosity. We check this by using the data of 
70 low-redshift radio-quiet 
quasars in the Palomar-Green survey. The emission line properties of these
quasars have been studied by Boroson \& Green (1992). The
continuum luminosity at 5100$\AA$ and the $H_\beta$ luminosity were estimated
from the absolute
V-band magnitude ($M_V$) and the equivalent width of $H_\beta$ emission line 
listed in their Table 1 and Table 2. Using the
$R-L_{H_\beta}$ and $R-L_{5100\AA}$ relations we  estimated the BLR sizes 
and black hole masses of these radio-quiet quasars. 
The results from the two relations are almost identical (see Figure 4).   
This is also indicated by the normalized 
$\chi^2$ value of the deviation of 
the points plotted in the upper panel of Figure 4, which is 0.94, much smaller than 
the value 2.58 for the points plotted in the upper panel of Figure 2 for 
radio-loud AGNs. From the lower panel of Figure 4, we can also see that the
difference of the two black hole mass estimates does not correlated with
 the radio-loudness for radio-quiet quasars. The Spearman's rank test 
gives a correlation
coefficient of only 0.05, much smaller than that for radio-loud AGNs. 

\begin{figure}
\includegraphics[width=9.5cm, height=12cm]{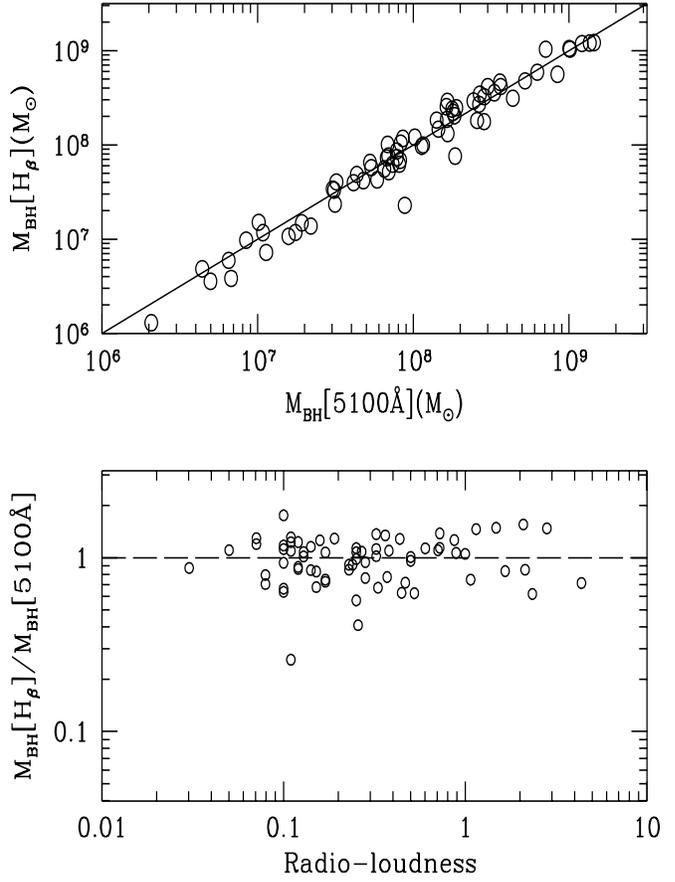}
 \caption{ Upper panel: Comparison of black hole masses of 70 radio-quiet 
PG quasars 
estimated with the $R-L_{H_\beta}$ and 
$R-L_{5100\AA}$ relations.  The diagonal
line shows the case where  the masses are identical. Lower panel: 
The  ratios of black hole masses estimated with two different 
relations are plotted against
 the radio-loudness of radio-quiet quasars.  The dashed line indicates 
that the case where the
two black hole mass estimates are identical.}
\end{figure}

\section{Discussions}

Using the empirical relation between the BLR 
size and optical 
continuum luminosity  possibly induces an overestimation of the BLR size 
and hence the 
black hole mass of  some extremely radio-loud
AGNs because of the jet contribution to the optical luminosity.
We derived another empirical relation between the BLR size and the emission 
line luminosity, and demonstrated that it can be used to estimate
the BLR size and black hole mass of both radio-quiet and radio-loud AGNs. 
 If the relativistic jets and host galaxy have significant contributions to 
the optical continuum, 
the emission line luminosity is probably a better tracer of ionizing
luminosity.
Comparisons of the estimated black hole masses with these two different
empirical relations clearly indicate that the difference becomes significant if
the radio-loudness of AGNs is larger. Using the  $R-L_{H_\beta}$ relation 
 may result in   more accurate estimations of black 
hole masses  of  some blazar-type AGNs.
 
 In our study we focused on the possible effects of beaming on  the optical continuum in 
radio-loud AGNs and ignored the difference in BLR physics between radio-loud and 
radio-quiet AGNs. The modest correlation between the
EW(H$_{\beta}$) and the radio-loudness  may indicate the presence of beaming effects, 
though some other effects such as a lower covering factor of the BLR of radio-loud AGNs can 
also lead to smaller EW (H$_{\beta}$) values.  Because
currently we know little about the difference of BLR physics between radio-loud and 
radio-quiet AGNs, to prove the validity of our approach it is necessary to  
compare 
 our estimated black hole mass with an independent estimate, for example,  
from the correlations of black hole mass with central velocity dispersion and host 
galaxy luminosity.
Unfortunately, not many measured values of central velocity dispersion or host galaxy
luminosity for extremely radio-loud AGNs are available. Although there are 8 objects 
in the sample of
Brotherton (1996) with measured host magnitude (McLure \& Dunlop 2001), the radio-loudness
of these objects are mostly smaller than 1000 and thus the difference estimated with
the $R-L_{H_\beta}$ and $R-L_{5100\AA}$ relations is rather small. The velocity 
dispersion measurements for radio-loud quasars are not available and the [OIII] profile in 
radio-loud AGNs may not be adopted to estimate the central velocity 
dispersion because of its complexity. Therefore, further imaging studies 
on the host galaxy and spectroscopic measurements of the central velocity
dispersions of a large sample of extremely radio-loud quasars are still 
desired to confirm our results.

The advantage of using the  $R-L_{H_\beta}$ relation is that we can estimate the 
black hole mass of AGNs with only two observed parameters, namely the 
H$_\beta$ line luminosity and its
FWHM, and it can be applied to a larger sample of AGNs with redshift smaller than 0.8. 
In principle, one can analogously investigate the relation between
the BLR size and the luminosity of some ultraviolet emission lines such as MgII and CIV, 
which may be used to estimate the black hole mass of some
high redshift AGNs. Some recent studies have suggested to use the ultraviolet continuum
luminosity and the FWHM of ultraviolet emission lines to estimate the black hole
mass of high redshift AGNs (Vestergaard 2002; McLure \& Jarvis 2002). However, the ultraviolet
continuum luminosity can similarly suffer the serious contaminations from jet and Blamer 
continuum, 
therefore the luminosity of  ultraviolet emission line again may be a better indicator of ionizing
luminosity than the  ultraviolet continuum luminosity.

Finally, one should be cautious to the uncertainties in estimating the black hole 
mass of AGNs using the  $R-L_{H_\beta}$ relation. Firstly, the variations of H$_\beta$ emission
line flux and its FWHM are common in AGNs. Estimating the black hole mass with the values of 
these two parameters in a single 
spectrum may lead to large errors. Secondly, the different
inclination of the BLR may also significantly affect the results (McLure \& Dunlop 2001; 
Wu \& Han 2001).  If the BLR has a flatten geometry and the inclination of BLR is rather small,
our derived values of black hole mass may significantly underestimated. However, the 
ratio of black hole masses estimated with two empirical relations
does not depend on the inclination. 
Better understandings of BLR 
geometry and dynamics are absolutely needed to diminish the uncertainties in deriving the 
black hole mass of AGNs (Krolik 2001).     

\begin{acknowledgements}
We thank Xinwu Cao and Dongrong Jiang for helpful discussions,  and the anonymous 
referee for valuable
suggestions which improve the paper significantly. The work is supported by
the National Key Project on Fundamental Researches (TG 1999075403),
the National Natural Science Foundation (No. 10173001) in China and the Jun Zheng
Foundation of Peking University.
This research has made use of the NASA/IPAC Extragalactic Database (NED) which is 
operated by the Jet Propulsion Laboratory, California Institute of Technology, 
under contract with the National Aeronautics and Space Administration. 
\end{acknowledgements}

\newpage
\tabcolsep 3.mm
  \begin{table*}
      \caption{The BLR size and luminosity data of 34 AGNs in the reverberation mapping studies}
         \begin{tabular}{lcccccc}
           \hline

Name    &   z    & $A_B$ &$R_{BLR}$	  & $ L_{5100\AA}$ &  $L_{H_\beta}$ & Ref\\
        &           & & (light-days)  &   ($10^{42}erg/s$) & ($10^{42}erg/s$) & \\
\hline
3C 120  &   0.033 & 0.570 & 42.0$^{+27.0}_{-20.0}$  &  73.00   $\pm$  13.00 &   1.222 $\pm$    0.120  & 1  \\
3C 390.3&   0.056 & 0.170 &   22.9$^{+    6.3}_{-    8.0}$  &   64.00  $\pm$   11.00 &  1.433   $\pm$   0.171 & 3  \\
Akn 120 &   0.033 & 0.400 &   37.4$^{+    5.1}_{-    6.3}$    &  139.00   $\pm$  26.00 &  2.496   $\pm$   0.463 & 1\\
F 9      &  0.046&  0.000 &  16.3$^{+    3.3}_{-    7.6}$    &  137.00   $\pm$  15.00&  2.056  $\pm$    0.174  & 4 \\
IC 4329A&   0.016&  0.000 &    1.4$^{+    3.3}_{-    2.9}$   &   16.40   $\pm$   2.10 &  0.183   $\pm$   0.014 & 5 \\
Mrk 110  &  0.035 & 0.000 &   18.8$^{+    6.3}_{-   6.6}$   &   38.00   $\pm$  13.00 &  0.811   $\pm$   0.203 & 1\\
Mrk 279 &   0.030 & 0.000 &   16.2$^{+    5.1}_{-    5.4}$   &   66.00    $\pm$  6.30 &  1.023   $\pm$   0.071 & 12\\
Mrk 335 &   0.026&  0.100&    16.4$^{+    5.1}_{-    3.2}$  &   62.20    $\pm$  5.70 &   1.160   $\pm$   0.073 & 1\\
Mrk 509 &   0.034&  0.180  &   76.7$^{+    6.3}_{-    6.0}$  &  147.00    $\pm$ 15.00 & 3.070   $\pm$   0.332  & 1\\
Mrk 590 &   0.026&  0.050 &   20.0$^{+    4.4}_{-    2.9}$  &  51.00    $\pm$  9.60 &   0.498   $\pm$   0.143  & 1 \\
Mrk 79  &   0.022&  0.230 &    17.7$^{+    4.8}_{-    8.4}$  &  42.30   $\pm$   5.60 &   0.619   $\pm$   0.041 & 1 \\
Mrk 817 &   0.031 & 0.000 &   15.0$^{+    4.2}_{-    3.4}$  &   52.60    $\pm$  7.70 &   0.740   $\pm$   0.123 & 1 \\
NGC 3227&   0.004 & 0.020 &     10.9$^{+    5.6}_{-  10.9}$  &    2.02    $\pm$  0.11 & 0.017    $\pm$  0.002  & 6\\
NGC 3783&   0.010 & 0.470 &     4.5$^{+    3.6}_{-   3.1}$   &   17.70    $\pm$  1.50&   0.292    $\pm$  0.021 & 7 \\
NGC 4051&   0.002 & 0.000 &    6.5$^{+    6.6}_{-    4.1}$   &   0.525    $\pm$  0.03 &   0.0048   $\pm$   0.0005  & 8\\
NGC 4151&   0.003 & 0.000 &     3.0$^{+    1.8}_{-    1.4}$   &   7.20    $\pm$  0.42 & 0.152    $\pm$  0.010  & 9 \\
NGC 5548 &  0.017 & 0.000 &   21.2$^{+    2.4}_{-    0.7}$   &  27.00    $\pm$  5.30 &   0.421   $\pm$   0.092  & 10\\
NGC 7469 &  0.016 & 0.120 &     4.9$^{+    0.6}_{-    1.1}$ &   55.30    $\pm$  1.60 &  0.423   $\pm$   0.019   & 11\\
PG 0026 &   0.142 & 0.130 &   113.0$^{+   18.0}_{-   21.0}$   &  700.00   $\pm$ 100.00 &  5.693   $\pm$   0.493 & 2\\
PG 0052 &   0.155 & 0.120 &   134.0$^{+   31.0}_{-   23.0}$  &  650.00   $\pm$ 110.00 & 9.595   $\pm$   1.119  & 2\\
PG 0804&    0.100 & 0.110 &   156.0$^{+   15.0}_{-  13.0}$  &  660.00   $\pm$ 120.00 & 13.95   $\pm$   0.908  & 2\\
PG 0844 &   0.064 & 0.080 &    24.2$^{+   10.0}_{-    9.1}$  &  172.00   $\pm$  17.00 & 2.585    $\pm$  0.257 & 2\\  
PG 0953&    0.239 & 0.000 &   151.0$^{+   22.0}_{-   27.0}$ & 1190.00   $\pm$ 160.00 &  19.39   $\pm$  1.129  & 2\\
PG 1211 &   0.085 & 0.130 &   101.0$^{+   23.0}_{-   29.0}$  &   493.00    $\pm$ 80.00 & 8.588    $\pm$  1.056 & 2\\
PG 1226 &   0.158 & 0.000 &  387.0$^{+   58.0}_{-   50.0}$ &  6440.00   $\pm$ 770.00&  88.37   $\pm$   7.192  & 2\\
PG 1229 &   0.064 & 0.000 &   50.0$^{+   24.0}_{-  23.0}$  &   94.00    $\pm$ 10.0 &   1.601    $\pm$  0.202  & 2\\
PG 1307 &   0.155 & 0.020 &   124.0$^{+   45.0}_{-   80.0}$  &  527.00    $\pm$ 52.00 &  9.603   $\pm$   1.301 & 2\\
PG 1411 &   0.089 & 0.000 &  102.0$^{+   38.0}_{-   37.0}$  &  325.00    $\pm$ 28.00 &  5.268   $\pm$   0.285  & 2 \\
PG 1426 &   0.086 & 0.120 &   95.0$^{+   31.0}_{-   39.0}$    &  409.00    $\pm$ 63.00 & 4.952   $\pm$   0.465  & 2\\
PG 1613 &   0.129 & 0.040 &   39.0$^{+   20.0}_{-   14.0}$   &  696.00    $\pm$ 87.00 &  7.014   $\pm$   0.451  & 2\\
PG 1617 &   0.114 & 0.150 &  85.0$^{+   19.0}_{-   25.0}$   &  237.00    $\pm$ 41.00 &  4.060   $\pm$   0.493  & 2 \\
PG 1700 &   0.292 & 0.020 &  88.0$^{+  190.0}_{-  182.0}$   & 2710.00   $\pm$ 190.00 &  35.29    $\pm$  1.851  & 2\\
PG 1704&    0.371 & 0.000 &  319.0$^{+  184.0}_{-  285.0}$  & 3560.00   $\pm$ 520.00 &  9.752    $\pm$  1.257  & 2\\
PG 2130 &   0.061 & 0.170 &  200.0$^{+   67.0}_{-   18.0}$    & 216.00   $\pm$  20.00 &   4.241    $\pm$  0.381 & 2 \\

            \hline
         \end{tabular}
\vskip 6mm
\noindent Notes: The Galactic extinction values are adopted from NED. 
Data of $R_{BLR}$ and $ L_{5100\AA}$ are taken from Kaspi et al. 
(2000). The H$_\beta$ luminosity is calculated from the flux given
in the literature (see the column ``ref''). References: (1) Peterson \etal 1998;
(2) Kaspi \etal 2000;
(3) Dietrich \etal (1998); (4) Santos-Lleo \etal (1997); (5) Winge 
\etal (1996); (6) Winge \etal (1995);
(7) Strirpe \etal (1994); (8) Peterson \etal (2000); (9) Kaspi \etal (1996);
(10) Peterson \etal (2002);
(11) Collier \etal (1996);
(12) Santos-Lleo \etal (2001). 

   \end{table*}

\end{document}